\documentclass[aps,pra,twocolumn,noshowpacs,superscriptaddress,floatfix]{revtex4-2}

\usepackage{graphicx}% Include ure files
\usepackage{dcolumn}% Align table columns on decimal point
\usepackage{bm}% bold math
\usepackage{mathtools}
\usepackage{xcolor}
\usepackage{layouts}
\usepackage{hyperref}
\usepackage{amssymb}
\usepackage{dsfont}
\usepackage{caption}
\usepackage{subcaption}

\usepackage{lmodern}
\usepackage{tikz}
\usetikzlibrary{decorations.pathmorphing}
\usepackage{lipsum} 

\usepackage[column=0]{cellspace}
\usepackage{blindtext}

\newcommand{\affilLAN}{Department of Physics, Lancaster University, Lancaster LA1 4YB, United Kingdom.}

\begin{document}

\tikzset{snake it/.style={decorate, decoration=snake}}

%\preprint{APS/123-QED}

\title{Stochastic action for the entanglement of a noisy monitored two-qubit system}

\author{Dominic Shea}\affiliation{\affilLAN}
\author{Alessandro Romito}\affiliation{\affilLAN}

\date{\today}

\begin{abstract}
    We study the effect of local unitary noise on the entanglement evolution of a two-qubit system subject to local monitoring and inter-qubit coupling. We construct a stochastic Hamiltonian by incorporating the noise into the Chantasri–Dressel–Jordan path integral and use it to identify the optimal entanglement dynamics and to develop a diagrammatic method for a closed-form approximation of the average entanglement dynamics with an analytical dependence on the noise and measurement intensity. We find that both the optimal trajectory and diagrammatic expansion capture the oscillations of entanglement at short times. Numerical investigation of long-time steady-state entanglement reveals a non-monotonic relationship between concurrence and noise strength.
\end{abstract}

\maketitle

\section{Introduction}

The phenomenon of quantum entanglement distinguishes quantum systems from their classical counterparts, and it is a central resource in quantum information processing, enabling a variety of technological advances such as superdense coding, quantum teleportation and quantum error correction~\cite{nielsen2000quantum,zeng2018quantum}.
Entanglement is also key to understanding some properties of condensed matter systems, such as many-body localization~\cite{PhysRev.109.1492,PhysRevLett.78.2803,BASKO20061126,ABANIN2021168415,RevModPhys.91.021001}.
This becomes particularly evident in classifying many-body quantum states, where systems may be distinguished according to the scaling law followed by their entanglement entropy~\cite{PhysRevLett.96.110404,Page_1993,zeng2018quantum}. 
%Topologically ordered states such as spin liquids and fractional hall states exhibit an area scaling law with an additional topological component to their entanglement.Some properties of Quantum effects in condensed matter, such as the Kondo Effect and Many-body localisation, may be understood in terms of quantum entanglement. 
Recently, entanglement has been exploited to identify out-of-equilibrium many-body states resulting from the stochastic dynamics of many body systems subject to random unitary evolution and quantum monitoring~\cite{PhysRevB.98.205136,PhysRevB.99.224307,PhysRevX.9.031009,PhysRevX.7.031016,PhysRevB.100.064204}.
This has witnessed the emergence of entanglement scaling transitions from a volume to an area law in a many-body Zeno effect along with a broader set of Measurement-induced Phase Transitions (MiPTs)~\cite{Fisher_2023}. These transitions appear when tracking the entanglement along individual quantum trajectories correlated with the measurement readout (as opposed to measurement-averaged dynamics), and have been reported in recent pioneering experiments~\cite{Noel_2022,Koh_2023,2023}.

Nascent features of the many-body competition between unitary evolution and monitoring in the entanglement dynamics can be identified even in two-qubit systems, in which different strategies to create, monitor, and maintain quantum entanglement can be implemented~\cite{PhysRevB.67.241305,Moehring_2007,Hutchison_2009,PhysRevA.81.040301, Bernien_2013,PhysRevA.83.022311}. 
The experiments reported in Refs.~\onlinecite{PhysRevLett.112.170501,PhysRevX.6.041052} have enabled the tracking of individual quantum trajectories for two entangled qubits. This development paves the way for the study of entanglement along trajectories in dual qubit systems.
While analytical expressions for the full probability distribution of steady states can be obtained only in particular cases~\cite{PhysRevResearch.2.033512},
recent studies investigated the statistics of the entanglement properties for two qubits under half-parity Gaussian monitoring~\cite{PhysRevX.6.041052} and the use of feedback control techniques to manipulate the purity and entanglement of dual qubit systems~\cite{PhysRevA.91.012118,PhysRevA.102.022612}. 

In this work, we study the monitored unitary dynamics in the presence of a stochastic unitary component, which can have non-trivial effects in many-body MIPTs~\cite{Kalsi_2022,PhysRevB.105.144202, Schomerus_2022}.
We apply the Chantasri–Dressel–Jordan (CDJ) path-integral to a two-qubit system subject to a driving Hamiltonian, two local Gaussian measurements and external noise. 
The CDJ path integral is a path-integral formulation explicitly developed for quantum systems subject to continuous Gaussian measurement and a driving Hamiltonian~\cite{PhysRevA.88.042110, PhysRevA.92.032125,Murch_2013}, which is especially advantageous for investigating rare events - providing a unique insight into quantum measurement dynamics~\cite{PhysRevA.95.042126,PhysRevA.98.012141,PhysRevA.97.012118,PRXQuantum.3.010327,PhysRevX.6.041052}.
%Recent research has investigated the effect of measurement on noisy multi-qubit systems. In particular, where the effect size of noise and quasi-continuous monitoring may be quantified it is observed that the presence of measuring causes a non-monotonic effect \cite{}. 
%The Chantasri–Dressel–Jordan (CDJ) path integral method has been developed specifically for quantum systems subject to continuous Gaussian measurement and a driving Hamiltonian. This formulation of the dynamics of a quantum system, applied previously to single qubits and the Harmonic oscillator \cite{} may also be applied to study multi-qubit systems. 
%A version of stochastic path integrals is also useful for applying diagrammatic tools to simple monitored systems \cite{}.  
We utilise these theoretical tools to develop a stochastic action and use a diagrammatic method to obtain a closed-form approximation for the average entanglement dynamics with an analytical dependence on the noise and measurement intensity.
Comparing the results from theory with numerical simulation demonstrates that the perturbative diagrammatic expansion captures the entanglement dynamics at short times.
We further demonstrate numerically the existence of a non-monotonic dependence of the steady state average entanglement, which is a unique feature of entanglement dynamics along quantum trajectories and would not be observable in the system's average dynamics.

\section{The model: two-qubit noisy monitored dynamics}

\begin{figure}[ht]
    \centering
    \begin{subfigure}{0.5\textwidth}
    \caption{}
    \includegraphics[width=0.9\textwidth]{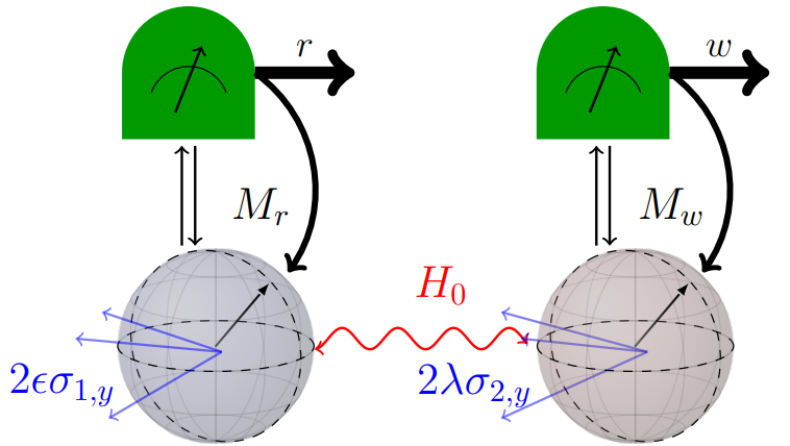}

    \end{subfigure}
    \begin{subfigure}{0.5\textwidth}
    \caption{}
    \includegraphics[width=0.9\linewidth]{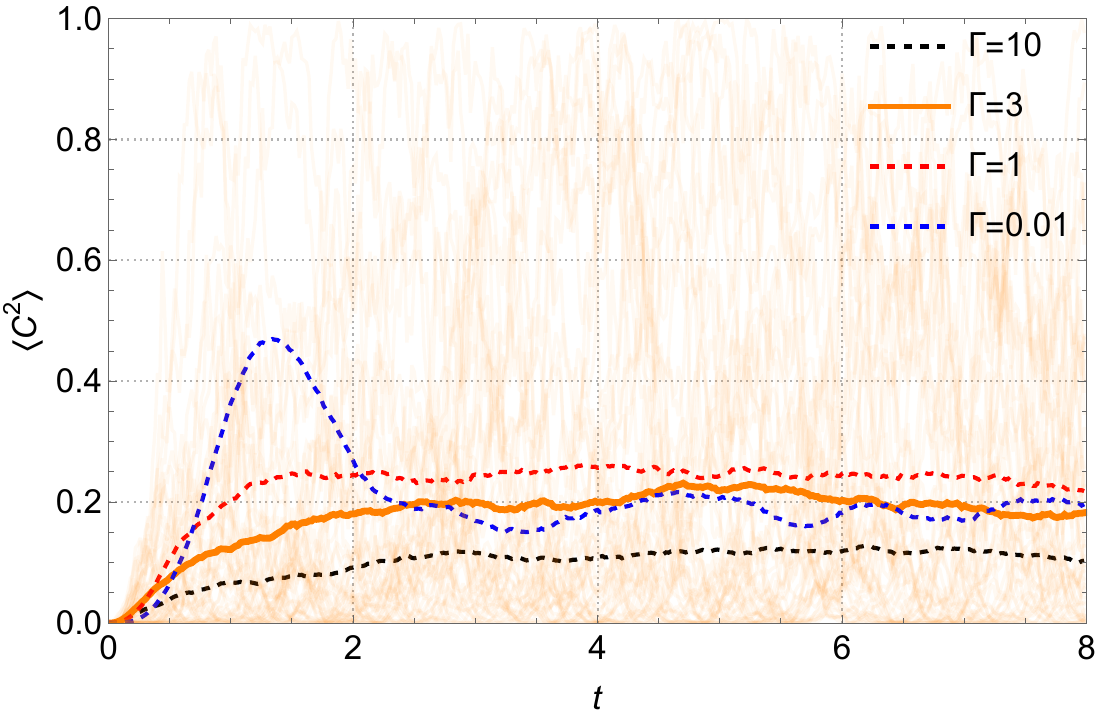}

    \end{subfigure}
    
    \caption{(a) Sketch of the two-qubit system studied in this paper. Two qubits (represented by single-qubit bock spheres) are coupled to quantum detectors (green) with output $r$, $w$ and Kraus operator backaction $M_r$, $M_w$. The qubit unitary dynamics is generated by an inter-qubit interaction term (red wavy line) and local noisy Hamiltonian terms (blue arrows). (b) Stochastic trajectories of $\mathcal{C}^2$  (fade orange background) from numerical simulations with fixed measurement strength ($\tau=0.2$) and noise strength ($\Gamma = 3 $) and corresponding average (orange full line). The average $\mathcal{C}^2$ vs time is plotted for a range of noise strengths at fixed measurement strength ($\tau = 0.2$). }
    \label{fig:cVSt}
\end{figure}

This paper considers a two-qubit system, represented schematically in Fig.~\ref{fig:cVSt}, subject to unitary noisy dynamics and continuous quantum monitoring. 
The unitary dynamics in the system are generated by the Hamiltonian $H$ specified as
\begin{align}
    H & = H_0 + H_n,\label{eq:Ham} \\
    H_0 & = i \hbar J (\sigma_{1,+}\sigma_{2,-}- \sigma_{1,-}\sigma_{2,+}), \label{eq:NoiseFreeHam} \\
    H_n &=  2 \epsilon(t) \sigma_{1,y} + 2 \lambda(t) \sigma_{2,y},
\end{align}
where $\sigma_{i,j}$ is the $j$-th Pauli matrix acting on the $i$-th qubit, $\sigma_{i,\pm} = \sigma_{i,x} \pm i \sigma_{i,y} $.
Fixing $\hbar =1$, and $J=1$ as an overall energy scale, $\epsilon$ and $\lambda$ become randomly fluctuating dimensionless Gaussian noises, defined by $\langle \epsilon(t) \epsilon(t') \rangle = \langle \lambda(t) \lambda(t') \rangle = \Gamma \delta(t-t')$. 
The parameter $\Gamma$ determines the noise strength, which we assume to be equal for both qubits.

The system is further subject to continuous Gaussian monitoring on each qubit~\cite{jacobs_2014,Jacobs_2006}, which is described by the state update over an infinitesimal time $\delta t$ as 
\begin{equation}
    \vert \psi_{t+\delta t}\rangle = \mathcal{N}^{-1} M_{1,r}\otimes M_{2,w} \vert \psi_{t}\rangle,
\end{equation} 
with the two Gaussian measurement operators acting on the first and second qubits, respectively given by
\begin{align}\label{eq:GaussianMeasurementOperators}
 M_r &= \sqrt[4]{\frac{\delta t}{2 \pi \tau}} \exp\left(-\frac{\delta t}{4 \tau}(r-\sigma_{1,z})^2 \right), \\  M_w &= \sqrt[4]{\frac{\delta t}{2 \pi \tau}} \exp\left(-\frac{\delta t}{4 \tau}(w-\sigma_{2,z})^2 \right), 
\end{align} 
and the normalization $\mathcal{N}^{-1}$ set by $\langle \psi_{t+\delta t}\vert \psi_{t+\delta t} \rangle=1$.
Here, the measurement readouts $r_k$ and $w_k$ are drawn from the probability distributions 
\begin{equation}
    P({r_k})=\langle \psi_t \vert M_{r_k}^\dagger M_{r_k} \vert \psi_t \rangle,
\end{equation}
\begin{equation}
    P(w_k)=\langle \psi_t \vert M_{w_k}^\dagger M_{w_k} \vert \psi_t \rangle.
\end{equation}
The measurement strength, $\tau$, is the time required to separate the readout distributions for $|0\rangle$ and $|1\rangle$ by two standard deviations~\cite{PhysRevA.92.032125}, where $|0\rangle$ and $|1\rangle$ correspond to a single qubit eigenstates of $\sigma_z$.
In the limit $\delta t \to 0$, this formulation is equivalent to the stochastic Schrodinger equation (SSE)~\cite{jacobs_2014}
\begin{align}
\label{eq:SSE}
    &d| \psi \rangle =\bigg[ -\frac{1}{\tau}(\sigma_{1,z}\sigma_{2,z}-\langle \sigma_{1,z}\sigma_{2,z} \rangle)^2 dt\\ \nonumber &+\sqrt{\frac{2}{\tau}}(\sigma_{1,z}-\langle \sigma_{1,z} \rangle) dW_1 +\sqrt{\frac{2}{\tau}}(\sigma_{1,z}-\langle \sigma_{1,z} \rangle) dW_2 \bigg] | \psi \rangle,
\end{align}
with Gaussian nose increments $dW_1$ and $dW_2$  so that $\langle dW_1 \rangle=\langle dW_2 \rangle = 0$ and variance $dW_1^2=dW_2^2 = dt$.

The evolution of the monitored system over a finite time $t$ is obtained as a succession of $N$ of these Kraus operators applied to the system so that $t = k \delta t$ with $\{ k \in \mathbb{N}| k \leq N \}$. 
The limiting procedure, defined by $N \to \infty, \delta t \to 0$ while keeping $\delta t \ N = t$, defines a continuous monitoring process as per the stochastic evolution of the SSE~\eqref{eq:SSE}. 
A continuous stream of measurement readouts is obtained: so $\{r_k\} \to r $ and $\{w_k\} \to w$, continuous random variables.
We can conveniently group all the stochastic variables of the model in the vector $\mathbf{s}= (r,w,\epsilon,\lambda)$. 
We parameterised our system with the state parameter $\mathbf{q}_k = (a_k,c_k,\alpha_k,\gamma_k)$ so that,  
\begin{equation}\label{eq:param}
    |\psi_k \rangle = a_k |00\rangle + c_k |01\rangle + \alpha_k |10\rangle + \gamma_k |11\rangle.
\end{equation}
After taking the continuous limit, these state parameters $\{\mathbf{q}_k\}$ also become continuous functions of time $\mathbf{q}(t)$.

We can formulate the time evolution from the combined quantum monitored and stochastic unitary evolution in a time-discretised formulation that is convenient for both the action formulation and the numerical implementation. 
In the discreet limit, the born rule gives the conditional probability of a pair of measurement readouts,
\begin{align}
\label{eq:bornrule}
\nonumber P(r_k,w_k| \psi_k, \lambda_k, \epsilon_k)&= \langle \psi_k  | E(r_k,w_k)^\dagger E(r_k,w_k) |\psi \rangle , \\
E(r_k,w_k)&= e^{-i H \delta t} (M_{r_k} \otimes M_{w_k}).
\end{align}
Here the unitary transformation $e^{-i H \delta t}$ is considered to act across the interval $[t,t+\delta t]$
with noise variables drawn from the independent Gaussian distributions 
\begin{align}\label{eq:noiseDistributions}
P(\epsilon_k) &= \sqrt{\frac{\delta t}{2 \pi \Gamma}} e^{-\frac{\delta t}{2 \Gamma}\epsilon_k^2},\\  P(\lambda_k)&=\sqrt{\frac{\delta t}{2 \pi \Gamma}} e^{-\frac{\delta t}{2 \Gamma}\lambda_k^2}, \end{align} across each time interval.
The effect of the Hamiltonian Eq.~\eqref{eq:Ham} and continuous monitoring is combined in the state update given by,
\begin{equation}\label{eq:genericpostmeasurementstate}
|\psi_{k+1} \rangle = E(r_k)|\psi_k \rangle /\sqrt{P(r_k, w_k)}.
\end{equation}
Taking this process back into the continuum limit produces dynamics that combine the effect of both the noisy Hamiltonian and the measurement (see Eqs.~(\ref{eq:SME}-\ref{eq:SME4coeff})). 

In the rest of the paper, we will study the stochastic dynamics of the qubit entanglement within this model. The time evolution is such that if the state preparation satisfies, $\mathbf{q}(0)\in \mathbb{R}$, then $\mathbf{q}(t)\in \mathbb{R}$ at all subsequent times. For the sake of simplicity, we study qubit dynamics with the initially unentangled preparation $|\psi(0)\rangle =(\frac{1}{2},\frac{1}{2},\frac{1}{2},\frac{1}{2})$. We do not expect the qualitative features of our analysis to be affected by this choice.

\section{Optimal entanglement}

We employ the concurrence, $\mathcal{C}$,
as an entanglement monotone to quantify the entanglement in the system. 
For the 2-qubit state parameterised in Eq.~\eqref{eq:param}, the concurrence is given by,
\begin{eqnarray}\label{eq:C}
    \mathcal{C} = 2 |a \gamma - \alpha c|.
\end{eqnarray}
The deterministic component of the Hamiltonian, $H_0$, can generate entanglement between the two qubits. 
If the system is initialised in a separable state, $H_0$ drives oscillations of entanglement with a period of $t =\pi $.
The local random Hamiltonians generically scramble these oscillations, leading to a stochastic evolution with a well-defined steady-state distribution of states and, hence, entanglement.
The local monitoring, a stochastic process, also drives the system towards a well-defined steady-state entanglement distribution. However, the entanglement dynamics are controlled by the competition between the measurements, which reduce the entanglement, and the unitary dynamics, which typically establish it. 

The stochastic entanglement evolution is reported in Fig.~\ref{fig:cVSt}, showing the average squared concurrence as a function of time for a range of different noise strengths.
Unless otherwise stated, results obtained from numerical simulations use the parameter $\delta t = 0.02$ and are based on a sample of size $400$ trajectories for any given value of the protocol parameters.
The average entanglement generically displays oscillations at short times before reaching a stated state value, which depends on the noise and measurement strength.

\subsection{Chantasri-Dressel-Jordan Path Integral with unitary noise}

We construct a path integral formulation of the probability distribution over quantum trajectories up to time $T$, $\mathcal{P}(\mathbf{q}(T))$ via the CDJ path integral~\cite{PhysRevA.92.032125} to identify the optimal entanglement dynamics.
The CDJ path integral is constructed by noting that the full probability distribution for the measurement process can be expressed as a product of sequential conditional probabilities so 
\begin{align}\label{eq:basic_CDJ}
&\mathcal{P}(\mathbf{q}(T), \mathbf{s}(T)|\mathbf{q}_i) = \mathcal{F}(t_0,t_N) \times \\ \nonumber  &\lim_{\delta t \to 0, N \to \infty} \prod_{k = 0}^{N - 1} P(\mathbf{q}_{k+1} | \mathbf{q}_k, \mathbf{s}_k) P(\mathbf{s}_k | \mathbf{q}_k),
\end{align}
where $\mathcal{F}(t_0,t_N)=\delta^3(\mathbf{q}(N)-\mathbf{q}_f) \delta^3(\mathbf{q}(0)-\mathbf{q}_i)$ sets the initial and final states to be $\mathbf{q}_i$ and $\mathbf{q}_f$. 

The path integral Eq.~\eqref{eq:basic_CDJ} therefore requires two critical pieces of information: the infinitesimal state update $P(\mathbf{q}_{k+1} | \mathbf{q}_k, \mathbf{s}_k)$ and the conditional measurement readout probability distribution $P(\mathbf{s}_k | \mathbf{q}_k)$.
The state updating term in Eq.~\eqref{eq:basic_CDJ} is deterministic, so it can be expressed as a delta function, 
\begin{align}\label{eq:diracdelta}
 P(\mathbf{q}_{k+1} | \mathbf{q}_k, \mathbf{s}_k)&= \delta^d(\mathbf{q}_{k+1}| \mathbf{q}_k, \mathbf{s}(t)) \\ \nonumber &=\Big(\frac{1}{2\pi i}\Big)^d \int^{i \infty}_{-i \infty} d^d \mathbf{p} e^{-\mathbf{p}_p\cdot(\mathbf{q}_k)}.
\end{align}
This step introduces conjugate momenta $\mathbf{p}_k=(p_a,p_c,p_\alpha,p_\gamma)$. Expanding both sides of Eq.~\eqref{eq:genericpostmeasurementstate} up to the first order in $\delta t$, we obtain, in the continuum limit, the stochastic differential equations for the evolution of the state parameters,
\begin{align}\label{eq:SME}
 \dot{a} &= A  -c \lambda-\alpha \epsilon, \\
 \dot{c} &= C + \alpha +a \lambda-\gamma \epsilon,\\
 \dot{\alpha} &= D - c  +a \epsilon-\gamma \lambda,\\ \label{eq:SME4}
 \dot{\gamma} &= Y +c \epsilon+\alpha \lambda,
\end{align}
where
\begin{align} \label{eq:SME1coeff}
    A&=\frac{a w}{\Gamma } (c^2 + \gamma^2) +\frac{a r}{\tau } (\alpha ^2+\gamma^2), \\ C &= \frac{c w}{\Gamma}(c^2+\gamma^2 -1) + \frac{c r}{\tau}(\alpha^2+\gamma^2) ,\\ D &= \frac{\alpha w}{\Gamma }(c^2 +\gamma^2) +\frac{r \alpha}{\tau }( \alpha^2 + \gamma^2 -1), \\ \label{eq:SME4coeff} Y &=  \frac{\gamma w}{\Gamma }(c^2+\gamma^2 -1)+\frac{r \gamma }{\tau }(\alpha^2+\gamma^2-1). 
\end{align}
Eqs.~(\ref{eq:SME1coeff}-\ref{eq:SME4coeff}) contain the effect of the measurement on the state dynamics. In the absence of these terms, Eqs.~(\ref{eq:SME}-\ref{eq:SME4}), describe the effect of any unitary noise ${\lambda(t),\epsilon(t)}$ on the system dynamics. We observe the noise free  ($\lambda = 0, \epsilon = 0$) and measurement-free oscillations ($r = 0, w = 0$) given by $\dot{c} = \alpha$, $\dot{\alpha} = -c$ are contained within Eqs.~(\ref{eq:SME}-\ref{eq:SME4}) as expected.

The Probability, $P(\mathbf{s}_k | \mathbf{q}_k)$ may be expressed as the product $P(\epsilon_k)P(\lambda_k)P(r_k,w_k|\mathbf{q}_k)$ and the factor $P(\epsilon_k)P(\lambda_k)$ is defined in Eq.~\eqref{eq:noiseDistributions}.
Expanding Eq.~\eqref{eq:bornrule} to leading (first) order in time, we find that the state-dependent conditional probability density is,
\begin{align} \label{eq:ProbDensity1}
&P(r_k,w_k|\mathbf{q}_k)= \frac{\delta t}{2 \pi   \tau } e^{ -\frac{\delta t}{2 \tau } (r_k^2+  w_k^2-2  w_k \Bar{w}_k -2  r_k \Bar{r}_k+2)},
\end{align}
where the mean readouts of the detectors are $\Bar{w}_k=\left(2 c_k^2+2 \gamma_k^2-1\right)$, $\Bar{r}_k=\left(2 \alpha_k^2+2 \gamma_k^2-1\right)$. 
Following the usual procedure for constructing path integrals~\cite{kleinert2009path,altland_simons_2010,PhysRevA.92.032125,PhysRevA.92.032125,PhysRevX.6.041052}, where extraneous factors of $\delta t$ are absorbed into the functional measure, we express $\mathcal{P}$ in terms of an action $\mathcal{S}[\mathbf{q},\mathbf{p}, \mathbf{s}]$, 
\begin{equation}\label{eq:path}
    \mathcal{P}(\mathbf{q}(T), \mathbf{s}(T)|\mathbf{q}_i) \propto \int D\mathbf{q} D\mathbf{p} D\mathbf{s} e^{- S[\mathbf{q},\mathbf{p}, \mathbf{s}]}.
\end{equation}
The functional $S$, acting on the whole set of quantum trajectories is given by
\begin{equation}\label{eq:action}
S[\mathbf{q}, \mathbf{p},\mathbf{s}] = \int_0^T \Big(-\mathbf{p} \cdot \dot{\mathbf{q}} + \mathcal{H}\Big) dt.
\end{equation}
Here $\mathcal{H}$ is expressed as
\begin{align}\label{eq:StochasticHam}
     &\mathcal{H}=p_a A +p_c C+p_\alpha D +p_\gamma Y-\frac{\lambda^2+\epsilon^2}{2\Gamma }\\ \nonumber &+\frac{1}{2 \tau} \Bigg(r^2+  w^2-2  r \Bar{r} -2  w \Bar{w}+ 2\Bigg),
\end{align}
and is referred to as stochastic Hamiltonian~\cite{PhysRevA.92.032125,PhysRevA.92.032125,PhysRevX.6.041052,PhysRevA.97.012118}, in analogy with the Hamiltonian formulation of a classic system. 
From this formulation, we see that the conjugate momenta act as Legendre multipliers, enforcing constraints on the rate of change of the state parameters. 
We will utilise this action formulation to investigate the optimal entanglement growth in the system.

\subsection{Optimal Squared Concurrence}

Extremising the CDJ stochastic action (Eq.~\eqref{eq:action}) via Hamilton's equation's $-\partial \mathbf{q} \mathcal{H} = \dot{\mathbf{p}},\partial \mathbf{p} \mathcal{H} = \dot{\mathbf{q}},\partial \mathbf{r} \mathcal{H} = 0$ leads to equations of motion that specify quantum trajectories that are extremum points of the action. Solutions of Eq.~\eqref{eq:SME} combined with Eq. (\ref{eq:mostprob1} - \ref{eq:mostprob8}) therefore constitute either most probable, least probable or saddle point trajectories for any given boundary conditions, the nature of any given solution is most easily determined by evaluating the action and comparing the solution with a nearby quantum trajectory. So for any accessible initial and final boundary conditions, we can use these equations (Eqs.~\eqref{eq:action}, \eqref{eq:SME} and~\eqref{eq:mostprob1}-\eqref{eq:mostprob8}) to determine the most likely quantum trajectory.

The extremization of Eq.~\eqref{eq:action} is carried out explicitly, resulting in 

\begin{align}\label{eq:mostprob1}
    \lambda  &= -\Gamma  (-a p_c+c p_a+p_\alpha \gamma -p_\gamma \alpha ) \\
    \epsilon &= \Gamma  (a p_{\alpha}+c p_\gamma-p_a \alpha -p_c \gamma ) \\
    \nonumber r &= a p_a \alpha ^2+a p_a \gamma ^2+c p_c \alpha ^2+c p_c \gamma ^2+p_\alpha \alpha  \gamma ^2+p_\alpha \alpha ^3\\
    &-p_\alpha \alpha +p_\gamma \alpha ^2 \gamma +p_\gamma \gamma ^3-p_\gamma \gamma -2 \alpha ^2-2 \gamma ^2+1,\\ \label{eq:mostprob4}
    \nonumber w &= a c^2 p_a+a p_a \gamma ^2+c^2 p_\alpha \alpha +c^2 p_\gamma \gamma +c p_c \gamma ^2+c^3 p_c \\
    & -c p_c-2 c^2+p_\alpha \alpha  \gamma ^2+p_\gamma \gamma ^3-p_\gamma \gamma -2 \gamma ^2+1.
\end{align}
Eqs.~(\ref{eq:mostprob1}-\ref{eq:mostprob4})express each stochastic variable in terms of the state parameters and conjugate momentum. 
These, in turn, are determined by the evolution of the system state parameters obtained earlier (Eq.~\eqref{eq:SME}), along with four further expressions for the evolution of the conjugate momentum variables, given by
\begin{widetext}
\begin{align}
\label{eq:mostprob5} \dot{p_a} &=-\frac{1}{\tau }\Big(c^2 p_a w+p_a r \alpha ^2+p_a r \gamma ^2+p_a \gamma ^2 w+\tau  p_\alpha \epsilon +\tau  p_c \lambda \Big) \\ \dot{p_c}&=\frac{1}{\tau}\Bigg(-2 a c p_a w-2 c p_\alpha \alpha  w-2 c p_\gamma \gamma  w-3 c^2 p_c w+4 c w+ \tau  p_\alpha+\tau  p_a \lambda -p_c r \alpha ^2-p_c r \gamma ^2-p_c \gamma ^2 w+p_c w-\tau  p_\gamma \epsilon \Bigg) \\
    \dot{p_\alpha}&=\frac{1}{\tau}\Bigg(-2 a p_a r \alpha +c^2 (-p_\alpha) w-2 c p_c r \alpha -\tau  p_c-3 p_\alpha r \alpha ^2-p_\alpha r \gamma ^2+p_\alpha r-p_\alpha \gamma ^2 w+\tau  p_a \epsilon -2 p_\gamma r \alpha  \gamma -\tau  p_\gamma \lambda +4 r \alpha \Bigg) \\ \nonumber
    \dot{p_\gamma} &=\frac{1}{\tau}\Bigg(-2 a p_a r \gamma -2 a p_a \gamma  w-2 c p_c r \gamma -2 c p_c \gamma  w+c^2 (-p_\gamma) w-2 p_\alpha r \alpha  \gamma +\tau  p_\alpha \lambda -2 p_\alpha \alpha  \gamma  w-p_\gamma r \alpha ^2-3 p_\gamma r \gamma ^2+p_\gamma r\\ \label{eq:mostprob8}&-3 p_\gamma \gamma ^2 w+p_\gamma w+\tau  p_c \epsilon +4 r \gamma +4 \gamma  w \Bigg).
\end{align}
\end{widetext}
Eqs.~(\ref{eq:mostprob1}-\ref{eq:mostprob4}) can be substituted into the remaining equations of motion, resulting in an 8-dimensional system of ODEs in phase space. The state initialisation $a(0) = c(0) = \alpha(0) = \gamma(0) =  \frac{1}{2}$, may then be combined with a final time boundary condition (or equivalently, with a choice of initial momentum) to generate extrema quantum trajectories.
While the exact solutions obtained from Eqs.~(\ref{eq:mostprob1}-\ref{eq:mostprob8}) depend on the initial state, we expect that the ensuing set of optimal trajectories will exhibit qualitatively and quantitatively similar behaviour in terms of their dynamic entanglement properties for generic non-entangled initial states.

\begin{figure}
\begin{subfigure}{0.5\textwidth}
    \caption{}
    \includegraphics[width = 0.8 \linewidth]{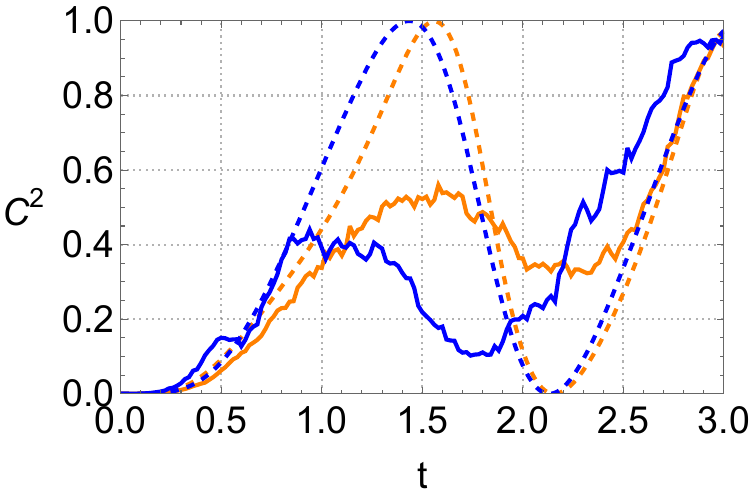}
\end{subfigure}
\begin{subfigure}{0.5\textwidth}
    \caption{}
    \includegraphics[width = 0.8 \linewidth]{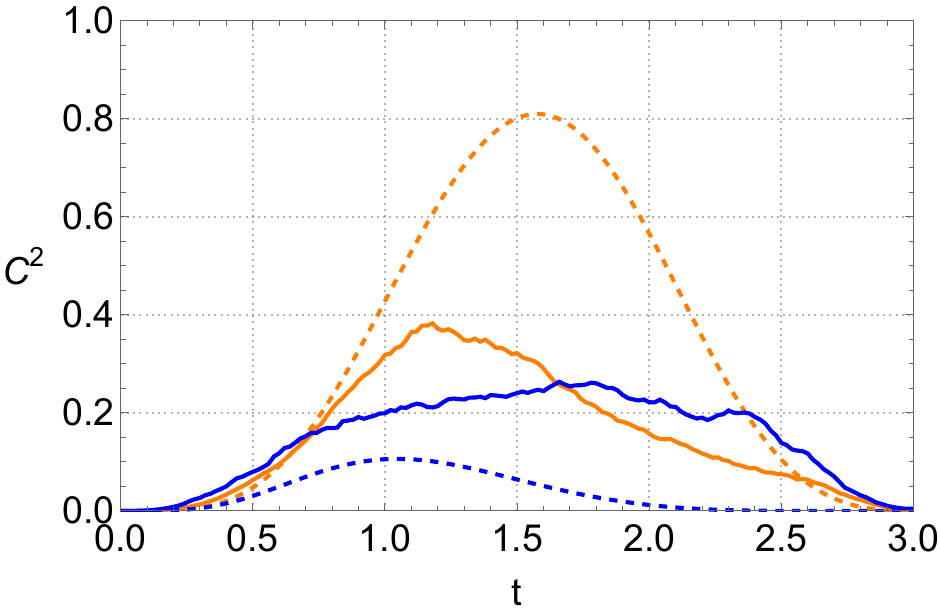}
\end{subfigure}
        
    \caption{Squared concurrence vs time for post-selected trajectories to maximum entanglement [panel (a)] and minimum entanglement [panel (b)] at $t=3$. Solid lines are for the average squared concurrence, and dashed lines are for the most probable trajectory with the specified boundary conditions for different noise strengths of $\Gamma=0.01$ (Orange) and $\Gamma=0.5$ (Blue). In all cases, $\tau = 0.2$. Across a range of noise strengths, an optimal concurrence track with a maximum concurrence at $t=\pi$ approximately reproduces the system's deterministic oscillations.   The optimal path with a final minimum entanglement is sensitive to changes in $\Gamma$. }
    \label{fig:optimum}
\end{figure}

To determine the most likely quantum trajectory for a given initial and final concurrence, we evaluate the probability of each candidate's optimal quantum trajectory over the equivalence class of final state conditions with equal entanglement monotone value (see Eq.~\eqref{eq:C}). Examples of these optimal entanglement trajectories are provided in Fig.~\ref{fig:optimum}, where they are compared to the associated post-selected average concurrence. We observe that the optimal entanglement captures some general features of the post-selected averages, particularly the number of inflexion points reproduced. However, these optimal curves are poor estimates of the average entanglement growth and attenuation.

\begin{figure}

        \includegraphics[width = 0.8 \linewidth]{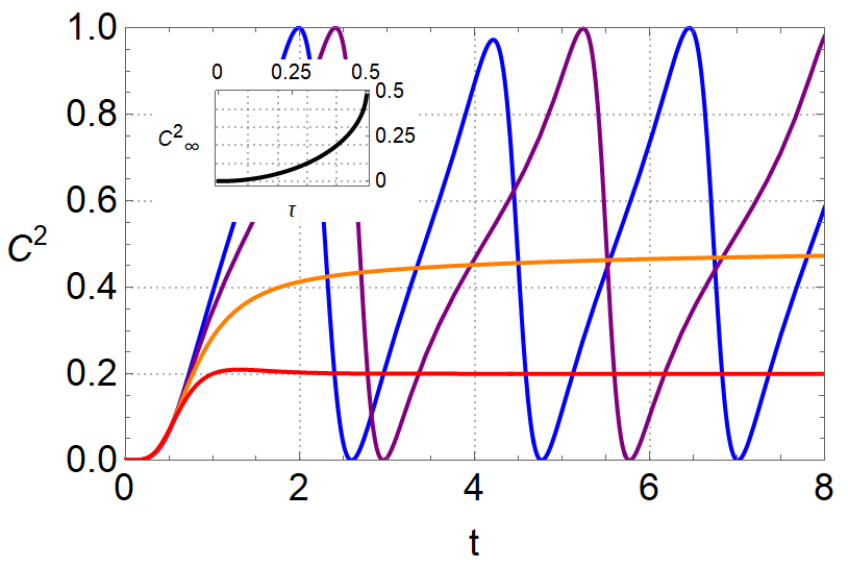}
    \caption{Global Optimum squared
    concurrence vs time for $\tau = 0.6$ (Purple), $\tau=0.7$ (Blue), $\tau = 0.4$ (Orange) and $\tau=0.3$ (Red). For $\tau>0.5$, $\mathcal{C}^2$ along the global optimal trajectory oscillates, while for $\tau<0.5$, it saturates to fixed values after a few units of time.  Inset: Plot of the steady-state global optimum concurrence against measurement strength for $\tau < 0.5$.}
    \label{fig:globalOpt}
\end{figure}

In addition to the most probable trajectories for given initial and final conditions, we may also identify the most probable concurrence globally (without any final boundary conditions). 
This is achieved by substituting in the values $\bar{w}= 1-2 c^2-2 \gamma^2, \bar{r} = 1-2 \alpha^2-2 \gamma^2, \lambda = 0, \epsilon = 0$ (the mean values of these stochastic variables at each time step, as can be identified from Eq.~\eqref{eq:ProbDensity1} into equations (\ref{eq:SME}-\ref{eq:SME4}).
The result is a set of ODEs that generated the global optimum trajectory.  
We see that the resulting equations of motion are then independent of $\Gamma$, hence the global optimum concurrence is independent of the noise strength. 
This is expected given that the unitary noise is independent of the system's state; hence, the noise fluctuations play no role.

We observe that the qualitative behaviour of the globally most probable entanglement undergoes a transition at $\tau \approx 0.5$.
Above $\tau \approx 0.5$, the globally optimum concurrence oscillates continuously as a function of time - dominated by the influence of the driving Hamiltonian.
The concurrence saturates to a fixed value below $\tau \approx 0.5$.
It is a monotonically decreasing function of the measurement strength, as might be expected in the strong measurement regime since measurement attenuates the entanglement.
This phenomenology is illustrated in Fig. \ref{fig:globalOpt}.
We note that the average entanglement also achieves a steady state - however, this occurs across all ranges of $\tau$ and generally saturates at a different value to the global optimum.

\section{Diagramatic Approximation for the average squared concurrence}

Besides the optimal behaviour, further analytical insight into the stochastic entanglement dynamics can be obtained for weak coupling/short-time dynamics, for which diagrammatic methods can be applied considering a suitable reformulation of the stochastic action~\cite{PhysRevA.92.032125}. 
We follow the procedure described in \cite{PhysRevA.92.032125,chow2012path} to construct a new stochastic action, again using Eq.~\eqref{eq:basic_CDJ} as the starting point. 
We will use a diagrammatic weak coupling expansion to find a closed-form expression for the average squared concurrence as a function of time that shows excellent agreement with the results of simulations in combined weak measurement weak noise regimes.

\subsection{ Ito rule formulation of the stochastic dynamics}

As a first step, we transform the state update equations (Eq.~\eqref{eq:SME}) into stochastic differential equations using Ito's rules, i.e. $ \delta t^2 r^2\to \delta t \tau ,\delta t^2 w^2\to \delta t \tau ,\delta t^2 \epsilon^2\to \Gamma  \delta t,\delta t^2 \lambda^2\to \Gamma  \delta t $. 
We apply these rules to the second-order expansion of the state update equation (Eq.~\eqref{eq:genericpostmeasurementstate}) to properly account for the noise fluctuations.  The measurement record functions are then approximated as their mean values plus some Gaussian noise, $r\to -\Bar{r} +\sqrt{\tau} \theta,w\to -\Bar{w}+\sqrt{\tau } \phi$. 
Here, $\theta$ and $\phi$ are new Gaussian noise variables with mean 0 and variance $\delta t^{-1}$. 
Additionally, we re-scale the earlier Gaussian noise variables with the transformation $\lambda \to \sqrt{\Gamma } \lambda,\epsilon\to \sqrt{\Gamma } \epsilon$
$\lambda$ $\epsilon$ so they also have a variance of $\delta t^{-1}$.
From this, we find the following SDEs governing the system state parameters
\begin{align}\label{eq:SDEs1}
 \nonumber \dot{a}&= -dt \bigg(\frac{a}{\tau} \tilde{A}+ a\Gamma \bigg)  +\frac{a \lambda}{\sqrt{\tau}}\big(\alpha^2 +\gamma^2\big)\\ &+\frac{a \lambda}{\sqrt{\tau}}(c^2 + \gamma^2 )-\sqrt{\Gamma} c \theta-\sqrt{\Gamma } \alpha \phi, \\ 
 \nonumber \dot{c} &=dt\bigg(\alpha -\Gamma  c +\frac{c}{\tau}\tilde{C}\bigg)+\frac{c \lambda}{\sqrt{\tau}}(\gamma^2+ c^2-1)\\&+\frac{c \epsilon}{\sqrt{\tau}}( \alpha^2+\gamma^2)-\sqrt{\Gamma } \gamma \phi +\sqrt{\Gamma} a \theta, \\
 \nonumber \dot{\alpha}&= dt\bigg(\frac{\alpha}{\tau}\tilde{D}-c-\Gamma  \alpha\bigg)+\frac{\alpha \lambda}{\sqrt{\tau}}( \gamma^2 + c^2)\\&+\frac{\alpha \epsilon}{\sqrt{\tau}}(\gamma^2 + \alpha^2 - 1)+\sqrt{\Gamma } a \phi-\sqrt{\Gamma } \gamma \theta, \\
\nonumber \dot{\gamma}&= dt\bigg(\frac{\gamma }{\tau} \tilde{Y}-\Gamma  \gamma\bigg)+\frac{\lambda}{\sqrt{\tau }}(c^2 \gamma +\gamma^3-\gamma)\\&+\frac{\epsilon}{\sqrt{\tau }}(\alpha^2 \gamma +\gamma^3-\gamma)+\sqrt{\Gamma } c \phi +\sqrt{\Gamma } \alpha \theta, \label{eq:SDEs4}
\end{align}
where
\begin{align*}
\tilde{A} &=\left(c^2 \gamma^2+\frac{c^4}{2}+\alpha^2 \gamma^2+\frac{\alpha^4}{2}+ \gamma^4\right),\\ 
\tilde{C} &=(-\alpha^2 \gamma^2 - \frac{\alpha^4}{2}-c^2 \gamma^2 + \gamma^2 -\gamma^4-\frac{c^4}{2}+c^2 -\frac{1}{2}),\\
\tilde{D} &=(-c^2 \gamma^2 - \frac{c^4}{2} -\alpha^2 \gamma^2 + \gamma^2 - \gamma^4 -\frac{\alpha^4}{2}+\alpha^2 -\frac{1}{2}), \\ 
\tilde{Y} &= (-c^2 \gamma^2 + c^2 -\frac{c^4}{2}-\alpha^2\gamma^2 + \alpha^2 -\frac{\alpha^4}{2} + \gamma^4+2\gamma^2-1).
\end{align*}
We observe that in the limit $\Gamma$ approaches zero, Eqs. (\ref{eq:SDEs1}-\ref{eq:SDEs4}) recover the stochastic Schrodinger equation~\eqref{eq:SSE}. 
These SDEs are utilized in the creation of a new path integral in the same manner as Eqs. (\ref{eq:SME}-\ref{eq:SME4}); each state parameter acquires a corresponding conjugate momentum through the Fourier representation of a Dirac delta function (Eq.~\eqref{eq:diracdelta}) in Eq.~\eqref{eq:basic_CDJ}.
We reuse the notation $\mathbf{p} = (p_a, p_c, p_\alpha, p_\gamma)$ to index these new conjugate momenta.

Each new Gaussian random variable, in discreet time, has an accompanying state-independent probability distribution,  $P(\theta) = \sqrt{\frac{\delta t}{2 \pi}}e^{-\frac{\delta t}{2} \theta^2} $ and $P(\phi) = \sqrt{\frac{\delta t}{2 \pi}}e^{-\frac{\delta t}{2} \phi^2} $.
Along with the transformed distributions from Eq.~\eqref{eq:noiseDistributions}, we multiply together these independent distributions to form the following probability density,
\begin{equation} \label{eq:newdistribution}
P(\theta,\phi,\lambda,\epsilon) = \frac{\delta t^2}{4\pi^2}e^{-\frac{\delta t}{2}(\theta^2+\phi^2+\epsilon^2+\lambda^2)} d \theta d\phi d\lambda d\epsilon.
\end{equation}
This stochastic action component is considerably simplified compared to its equivalent in Eq.~\eqref{eq:ProbDensity1}. 

A second stochastic path integral may be constructed from this new model of the system \cite{chow2012path,PhysRevA.92.032125} substituting in time discretised versions of the Fourier-transformed Dirac delta distributions associated with Eq.~\eqref{eq:SDEs1} and the distribution of Eq.~\eqref{eq:newdistribution} into Eq.~\eqref{eq:basic_CDJ}.
Taking this into the continuum limit and then integrating out all four Gaussian noises from the resulting path integral produces a phase space path integral
\begin{equation}\label{eq:phasespaceSPI}
    \mathcal{P} \propto \int \mathcal{D}\mathbf{q}\mathcal{D}\mathbf{p} e^{\tilde{S}}.
\end{equation}
The stochastic action $\tilde{S}$ is composed of two parts
\begin{equation}\label{eq:stochasticAction}
    \tilde{S} = \tilde{S}_f + \tilde{S}_i+\tilde{S}_b,
\end{equation}
a free Action, $\tilde{S}_f$, containing all bilinear terms of the action 
\begin{align}\label{eq:freeAction}
 \nonumber &\tilde{S}_f= \int p_a \dot{a}+ \Gamma  a p_a+ p_c \dot{c}+ c p_\alpha+\Gamma  c p_c+\frac{c p_c}{2 \tau }- p_c \alpha \\ &+p_{\alpha}\dot{\alpha} +\Gamma  p_{\alpha}\alpha+\frac{p_\alpha  \alpha}{2 \tau }+p_\gamma  \dot{\gamma}+\Gamma  p_\gamma  \gamma +\frac{p_\gamma \gamma }{\tau } dt.
\end{align}
The interaction action, $\tilde{S}_i$, containing all remaining terms
\begin{equation}
\label{eq:interaction-easy}
    \tilde{S}_i=\sum_{\rho} \mathcal{F}_{\mathcal{\rho}} a^{n_a}c^{n_c}\alpha^{n_\alpha}\gamma^{n_\gamma} p_a^{m_a} p_c^{m_c}p_\alpha^{m_\alpha}p_\gamma^{m_\gamma},
\end{equation}
where the sum is over all the strings $\rho=\{n_a,n_c,n_\alpha,n_\gamma,m_a,m_c,m_\alpha,m_\gamma \}$ such that $\sum_j n_j \leq 6$ and the coefficients of each terms are explicitly given in Appendix~\ref{ap:interaction}, Eq.~\eqref{eq:interactionAction}. 
The final term in the action expresses the boundary conditions 
\begin{equation}
    \tilde{S}_b=\frac{1}{2} \delta (t) p_a+\frac{1}{2} \delta (t) p_c+\frac{1}{2} \delta (t) p_\alpha+\frac{1}{2} \delta (t) p_\gamma.
\end{equation}
We used the state normalization condition, $a^2 + c^2 + \alpha^2 + \gamma^2 = 1$, to simplify the result.
One advantage of having the action in this form is that the free part of the action in Eq.~\eqref{eq:freeAction} is exactly solvable, and it becomes possible to apply various diagrammatic methods.
In the following section, we use this path integral to find a closed-form approximation for concurrence as a function of time.

\subsection{Diagrammatic Approximation for the squared Concurrence}

We aim to compute the moments of the squared concurrence $\langle \mathcal{C}^2\rangle = 4(\langle a^2 \gamma ^2 \rangle -2 \langle a c \alpha  \gamma \rangle +  \langle \alpha ^2 c^2 \rangle)  $.
To this end, various approximation techniques~\cite{chow2012path,wio2013path} may be applied to Eq.~\eqref{eq:phasespaceSPI}. 
We are interested specifically in a diagrammatic method for a weak coupling approximation~\cite{chow2012path} that gives a closed-form expression for the squared concurrence.  A weak coupling approximation involves expanding around the free action $\tilde{S}_f$, for which we have a closed-form expression for the Green's function $G$ associated with the free propagator,
\begin{equation}
\label{eq:propagator}
 G(t,t') = \Theta (\Delta t)  e^{\frac{-(2 \Gamma  \tau +1)\Delta t}{2 \tau }} P,
\end{equation}
where $\Theta(t)$ is the Heaviside function (with $H(0)=0$ due to Ito's condition and $\lim_{\Delta t \to 0^+}H(\Delta t) = 1$)
and  $P$ is a matrix of differential operators which act on a vector of four-dimensional real-valued functions $\boldsymbol{q}(t) = (q_1,q_2,q_3,q_4)=(a,c,\alpha,\gamma)$,
\begin{equation}
   P = \left(
\begin{array}{cccc}
  e^{-\Delta t(2 \Gamma +\frac{1}{2 \tau })} & 0 & 0 & 0 \\
 0 & \cos (\Delta t) &  \sin (\Delta t) & 0 \\
 0 &  -\sin (\Delta t) &  \cos (\Delta t) & 0 \\
 0 & 0 & 0 &  e^{-\Delta t (2 \Gamma +\frac{3}{2 \tau })} \\
\end{array} \right),
\end{equation}
with $ \Delta t = t - t'$. We may consider the $(i,l)$-th component of the matrix corresponding to pre-multiplication by the $p_l$ component ($\boldsymbol{p}(t) = (p_1,p_2,p_3,p_4)=(p_a,p_c,p_\alpha,p_\gamma)$), of momenta and post multiplication with the $q_j$ state parameter. This free propagator will be the basis to evaluate Feynman diagrams for averages of state-dependent quantities and expansion of the interaction action $\tilde{S}_i$.

We associate a Feynmann diagram with the average of a monomial in the state and momentum variables. 
Averages are performed over the free action and are computed via the green function in Eq.~\eqref{eq:propagator}.
Following the general construction in ~\cite{chow2012path}, 
each term in $\tilde{S}_i$ is associated with a single vertex in a Feynman diagram characterised by the edges it connects with. Different edges, labelled by different lines (full/dashed/dotted/wiggly), correspond to different state parameters as per the correspondence in Table~\ref{tab:one}. 
To fully specify the diagram, we further
associate ingoing arrows with state parameters and outgoing arrows with conjugate momenta.

\begin{figure}[ht]
    \centering
    \includegraphics[width=0.7\linewidth]{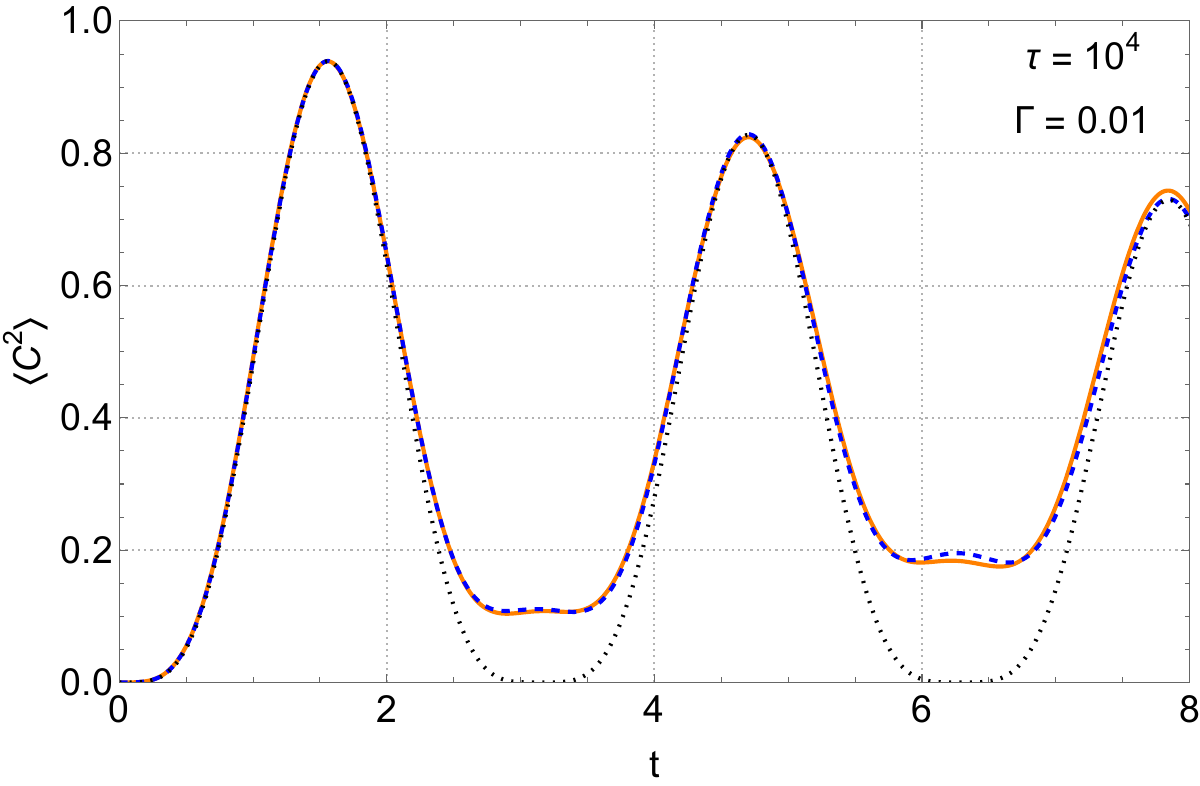}
    \includegraphics[width=0.7\linewidth]{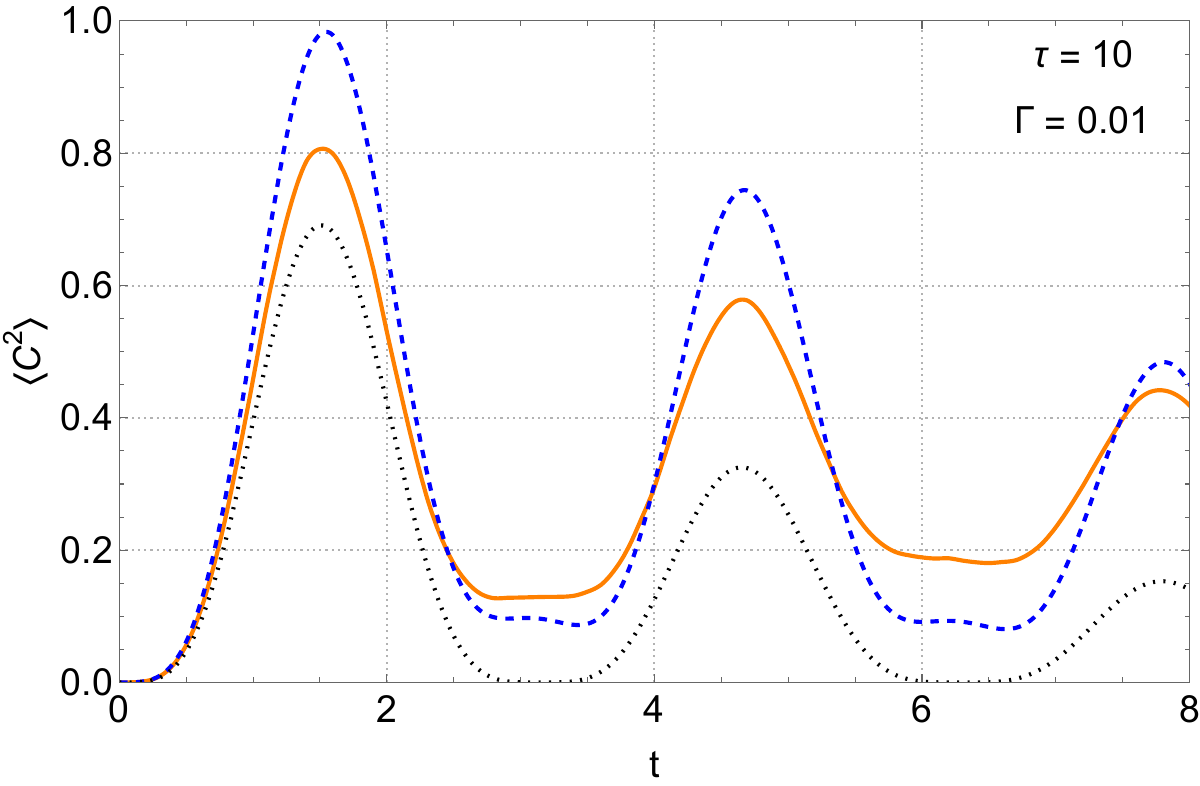}
    \includegraphics[width=0.7\linewidth]{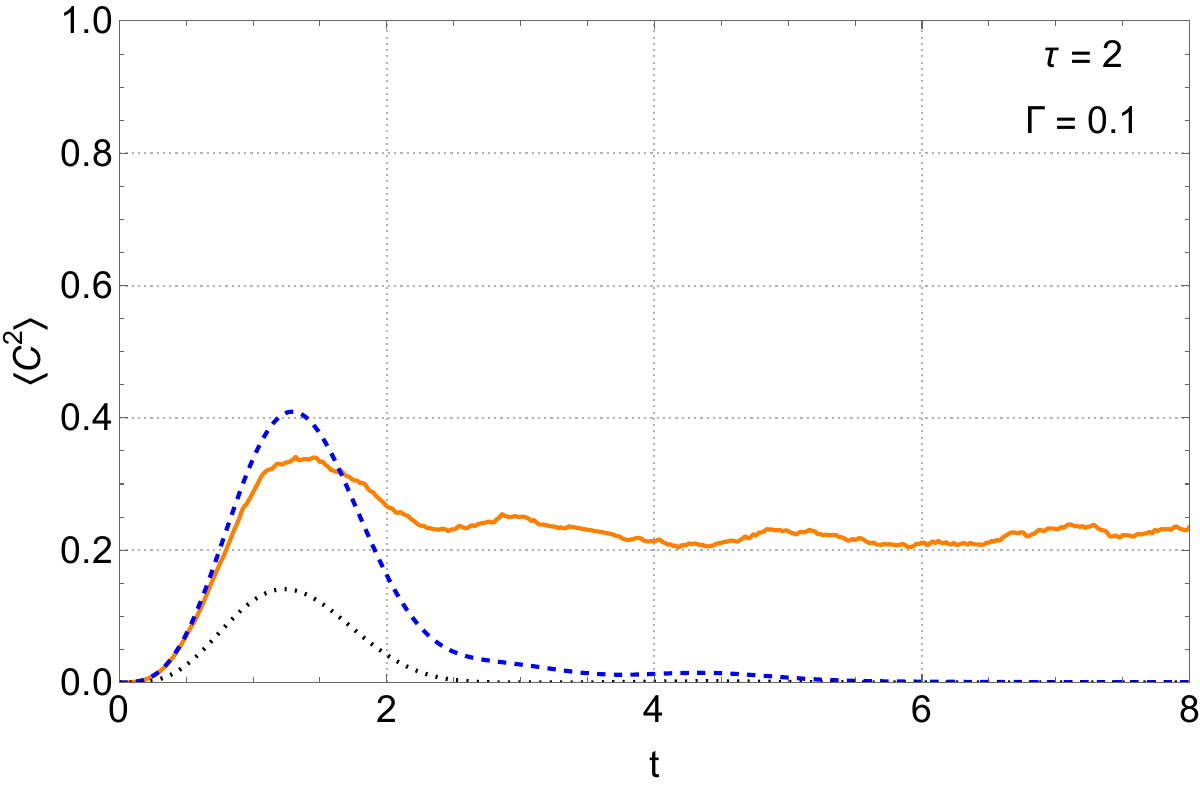}
    \caption{Average squared concurrence against time. Numerical simulations (solid orange lines) are compared with the linear approximation (dotted black) and the weak coupling approximation ---Eq.~\eqref{eq:diagram}--- (dashed blue) for different measurement and noise strengths. The weak coupling approximation outperforms the linear approximation and shows excellent agreement with numerical simulations in the weak measurement regime}
    \label{fig:Diagramatics}
\end{figure}

\begin{center}
\begin{table}[ht]
\begin{tabular}{|c|c|c|}
    \hline
    State Variable & Momentum & Edge  \\ [0.5ex] 
 \hline\hline
       a & $p_a$  & \begin{tikzpicture} 
       \draw[thick,-<] (0,0) -- (0.4,0);
  \draw[thick,-] (0.4,0) -- (0.8,0);\end{tikzpicture} \\
       c & $p_c$  & \begin{tikzpicture} 
       \draw[thick,dashed,-<] (0,0) -- (0.4,0);
  \draw[thick,dashed,-] (0.4,0) -- (0.8,0);\end{tikzpicture}  \\
      $\alpha$ & $p_\alpha $ & \begin{tikzpicture} 
       \draw[thick,dotted,-<] (0,0) -- (0.4,0);
  \draw[thick,dotted,-] (0.4,0) -- (0.8,0);\end{tikzpicture} \\
       $\gamma$ & $p_\gamma$ & \begin{tikzpicture} 
       \draw[thick,snake it,-<] (0,0) -- (0.4,0);
  \draw[thick,snake it,-] (0.4,0) -- (0.8,0);\end{tikzpicture} \\ \hline
    \end{tabular}
    \caption{Association between edge type and state variable type for the construction of Feynmann diagrams.}
    \label{tab:one}
    \end{table}
\end{center}
For example, the vertex associated with $\alpha \gamma p_a p_c$, which corresponds to the factor $ \Gamma \int_0^T p_a p_c \alpha \gamma dt $ in Eq.~\eqref{eq:interactionAction}, is given by the diagram in Fig.~\ref{fig:Diagramatics}(a).
%The type and number of ingoing edges have been determined by the state parameters $\alpha \gamma$ while the outgoing edges are specified outgoing conjugate momenta $p_a p_c$. 
Similarly, the vertex associated with $p_c c \gamma^2$ ($\int_0^T p_c c \gamma^2 dt$ in Eq.~\eqref{eq:interactionAction}) is shown in Fig.~\ref{fig:Diagramatics}(b). 
The vertex in Fig.~\ref{fig:Diagramatics}(a) consists of two state parameters and two conjugate momenta and the corresponding diagram features two ongoing and two outgoing edges; the vertex in Fig.~\ref{fig:Diagramatics}(b) instead includes three state parameters but only one momentum and its diagram accordingly requires three ingoing and one outgoing edge.
Finally, one outgoings diagrams for terms that include no state parameters, such as $p_a$, which corresponds to the term $\frac{1}{2} \int_0^T p_a \delta(t) dt $ in $\tilde{S}_b$. In those cases, the diagrams are simple, consisting of a single edge. 
As an example, the diagram for $\frac{1}{2} \int_0^T p_a \delta(t) dt $ is in Fig.~\ref{fig:Diagramatics}(c).

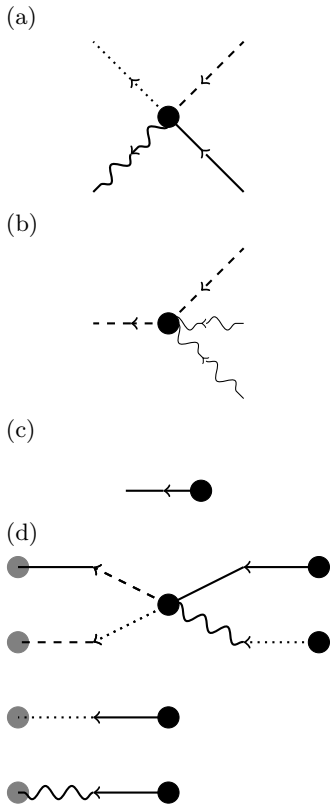
\begin{figure}

\begin{subfigure}{0.5\linewidth}
    \caption{}
    \centering

    \begin{tikzpicture}
    % Draw the black dot vertex
    \filldraw [black] (0,0) circle (4pt);
  
    % Draw the edges radiating outwards
    \draw[thick,dashed,-<] (0,0) -- (0.5,0.5);
    \draw[thick,dashed,-] (0.5,0.5) -- (1,1);
  
    \draw[thick,dotted,->] (0,0) -- (-0.5,0.5);
    \draw[thick,dotted,-] (-0.5,0.5) -- (-1,1);
  
    \draw[thick,snake it,->] (0,0) -- (-0.5,-0.5);
    \draw[thick,snake it,-] (-0.5,-0.5) -- (-1,-1);
    
     \draw[thick,-<] (0,0) -- (0.5,-0.5);
    \draw[thick,-] (0.5,-0.5) -- (1,-1);
\end{tikzpicture}
\end{subfigure}

\begin{subfigure}{0.5\linewidth}
\caption{}
    \begin{tikzpicture}
  % Draw the black dot vertex
  \filldraw [black] (0,0) circle (4pt);
  
  % Draw the edges radiating outwards
  \draw[thick,dashed,-<] (0,0) -- (0.5,0.5);
  \draw[thick,dashed,-] (0.5,0.5) -- (1,1);
  
  \draw[thick,dashed,->] (0,0) -- (-0.5,0);
  \draw[thick,dashed,-] (-0.5,0) -- (-1,0);

    \draw[snake it,-<] (0,0) -- (0.5,0);
  \draw[snake it,-] (0.5,0) -- (1,0);
    
  \draw[snake it,-<] (0,0) -- (0.5,-0.5);
  \draw[snake it,-] (0.5,-0.5) -- (1,-1);
\end{tikzpicture}
\end{subfigure}

\begin{subfigure}
{0.5\linewidth}
\caption{}
    \begin{center}
    \begin{tikzpicture}
  % Draw the black dot vertex
  \filldraw [black] (0,0) circle (4pt);
  
  % Draw the edges radiating outwards
  
  \draw[thick,->] (0,0) -- (-0.5,0);
  \draw[thick,-] (-0.5,0) -- (-1,0);
\end{tikzpicture}
\end{center}
\end{subfigure}

\begin{subfigure}{0.5\linewidth}

\caption{}
\centering
    \begin{tikzpicture}
    % Draw the gray dot vertex
    \filldraw [gray] (-4,3) circle (4pt);
    % Draw the gray dot vertex
    \filldraw [gray] (-4,2) circle (4pt);
    % Draw the gray dot vertex
    \filldraw [gray] (-4,1) circle (4pt);
    % Draw the gray dot vertex
    \filldraw [gray] (-4,0) circle (4pt);

     % Draw the edges
    \draw[thick,-] (-3,3) -- (-4,3);
    % Draw the edges
    \draw[thick,dashed,-] (-3,2) -- (-4,2);
    
    % Draw the edges
    \draw[thick,dotted,-] (-3,1) -- (-4,1);
    % Draw the edges
    \draw[thick,snake it,-] (-3,0) -- (-4,0);

     % Draw the edges
    \draw[thick,->] (-2,1) -- (-3,1);
    % Draw the edges
    \draw[thick,->] (-2,0) -- (-3,0);

    % Draw the edges radiating outwards
    \draw[thick,dashed,->] (-2,2.5) -- (-3,3);
    \draw[thick,dotted,->] (-2,2.5) -- (-3,2);
     % Draw the edges radiating outwards
    \draw[thick,dashed,->] (-2,2.5) -- (-3,3);
    \draw[thick,dotted,->] (-2,2.5) -- (-3,2);
    % Draw the black dot vertex
    \filldraw [black] (-2,2.5) circle (4pt);
    % Draw the edges radiating outwards
    \draw[thick,-] (-1,3) -- (-2,2.5);
    \draw[thick,snake it,->] (-1,2) -- (-2,2.5);

     \draw[thick,dotted,->] (0,2) -- (-1,2);
     \draw[thick,->] (0,3) -- (-1,3);
    
    % Draw the black dot vertex
    \filldraw [black] (-2,1) circle (4pt);
    % Draw the black dot vertex
    \filldraw [black] (-2,0) circle (4pt);
    % Draw the black dot vertex
    \filldraw [black] (0,3) circle (4pt);
    % Draw the black dot vertex
    \filldraw [black] (0,2) circle (4pt);

\end{tikzpicture}
\end{subfigure}

    \caption{(a)-(c) Examples of interaction vertices associated with terms in Eq.~\eqref{eq:interactionAction}. (d) An example of a Feynman diagram which contributes to the calculation of $\langle a c \alpha \gamma \rangle$. } 
    \label{fig:interactionvertex}
\end{figure}

When computing averages of state-dependent quantities perturbatively, the monomials to average will be included as edges of the Feynman diagrams. 
Specifically, the squared concurrence of interest here, $\langle C^2 \rangle = 4(\langle a^2 \gamma ^2 \rangle -2 \langle a c \alpha  \gamma \rangle +  \langle \alpha ^2 c^2 \rangle)$, consists of three terms, each with four state parameters. Each of the three terms may be approximated separately. 
For example, for the term $\langle ac\alpha\gamma\rangle$, one of Feynman's diagrams contributing to the average to order five IV is presented in Fig.~\ref{fig:Diagramatics}(d). The state parameters are associated with vertices (grey-coloured vertices in Fig. \ref{fig:interactionvertex}),  placed at the final time position in our Feynman Diagrams.
The remaining vertices (black) are from the expansion of the interaction action. 

Time-directed (right to left) Feynman diagrams may now be constructed by selecting all combinations of interaction vertices that can be connected to the appropriate final time vertices.
All diagrams contributing non-vanishingly to this approximation must connect with final vertices determined by the state parameters we are averaging, with no unconnected edges left over (see Fig. \ref{fig:interactionvertex}).
Note that since the quadratic action mixes terms with different state parameters and conjugates momenta, the propagators can connect different types of edges.

We sum up all such Feynman diagrams and then convert them into equations according to the following rules \cite{chow2012path}
\begin{enumerate}
    \item[1a.] For every interaction vertex in the diagram, a prefactor of $-\frac{1}{n!}$ is added. Where $n$ is the count of conjugate momenta in that vertex.
    \item[1b.] If the same vertex occurs $k$ times in a diagram, then there is an additional factor of $\frac{1}{k!}$.
    \item[1c.]  If there are $p$ distinct ways of connecting edges to vertices that yield the same diagram there is a final prefactor factor of $p$.
    \item[2.] Each edge between time $t_j$ and $t_k$  is replaced with the $q_i$-th $p_l$-th  component of the free propagator $G_{p_l,q_i}(t_j,t_k)$.
    \item[3.] Integrate from $0$ to $\infty$ over the time index associated with each interaction vertex.
\end{enumerate}

We organise our approximation by the number of Interaction Vertices (IV) used. We include terms up to five IVs to obtain a closed form expression for $\langle C^2 \rangle$ in Eq.~\eqref{eq:diagram}. Note that three or fewer IV diagrams leave unconnected edges, and using only four IVs is equivalent to only solving the linear component of Eq.~\eqref{eq:SDEs1}. This linear approximation is given by the first term in Eq.~\eqref{eq:diagram}, which corresponds to diagrams with four vertices, while subsequent terms are built from five IV.

\begin{align}
\label{eq:diagram}
\nonumber
\langle C^2 \rangle \approx & \sin^4(t) e^{-2 t(2 \Gamma +\frac{1}{\tau })} +\frac{1}{8 \tau } e^{-2 t \left(2 \Gamma +\frac{1}{\tau }\right)} \\
\nonumber &\Bigg(\frac{64 \Gamma  \tau ^4 \sinh \left(\frac{t}{\tau }\right)}{4 \tau ^2+1}-\frac{32 \Gamma  \tau ^3 \sin (2 t)}{4 \tau ^2+1} \\
\nonumber & -4 \Gamma  \tau  \sin (4 t)+2 t (8 \Gamma  \tau -5) \cos (2 t)\\ 
&+9 t+\sin (2 t)-\sin (4 t)+3 t \cos (4 t)\Bigg) 
\end{align}

We plot Eq.~\eqref{eq:diagram} for a selection of measurement and noise strength choices in Fig.~\ref{fig:Diagramatics}. 
The diagrammatic approximation is valid for a combination of weak noise and measurement strength. Eq.~\eqref{eq:diagram} outperforms the linear approximation, which consists of only the first term in Eq.~\eqref{eq:diagram}, (also plotted in Fig.~\ref{fig:Diagramatics}): the latter captures the oscillations with the correct frequency at short times, but misses the correct values of amplitudes. 
Notably, it is evident from Fig.~\ref{fig:Diagramatics} that, after a period of oscillating behaviour, the system's average entanglement will eventually enter a steady state (cf. also Fig. \ref{fig:CvsGamma}). 
We note that our weak coupling approximation at five IVs fails to capture this regime across all values of the measurement and noise strength; in fact, the long time limit of Eq.~\eqref{eq:diagram} corresponds to a vanishing entanglement. 

\section{Steady-state entanglement}
\begin{figure}
    \centering
    \includegraphics[width=0.9\linewidth]{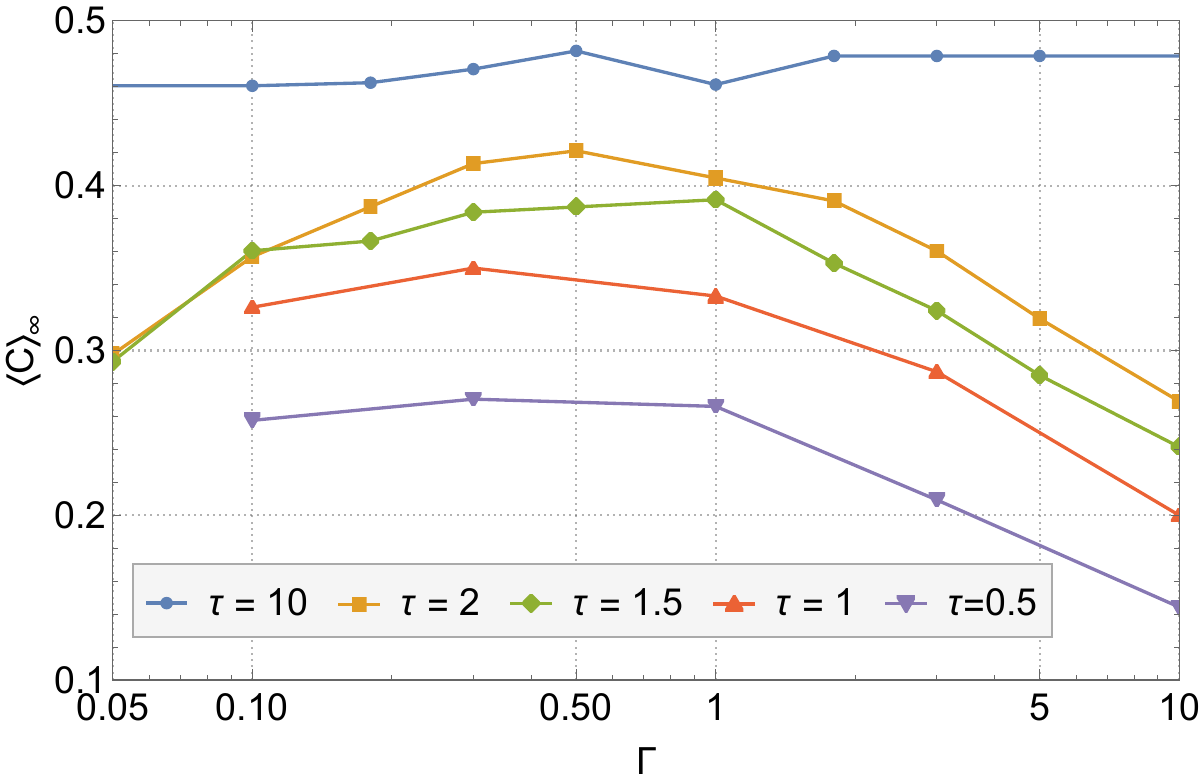}
    \caption{Steady state concurrence vs noise intensity, $\Gamma$, for a range of measurement strengths. Entanglement exhibits a non-monotonic dependence on noise strength.}
    \label{fig:CvsGamma}
\end{figure}

%Multi-qubit set-ups that feature both monitoring and noise, similar to the system we study, have been shown to attain a steady state average entanglement. 
While the diagrammatic approximation developed above can only capture the entanglement dynamics at short times and predicts vanishing steady-state concurrence,  the system is generically expected to reach an average steady-state entanglement from the competition of the unitary entangling terms and measurement disentangling effects. 
This competition is the source of measurement-induced entanglement transitions in many-body systems, and a steady-state interplay is already present at the level of 2 qubits.
We study the steady state of the system numerically.
The results are reported in Fig. \ref{fig:CvsGamma} for both entanglement monotones of interest. 

In the absence of measurements, the combination of unitary dynamics and Gaussian noise is expected to lead to a uniformly distributed `ergodic' steady state, where every accessible final state is realised with equal probability.
Using the values of $\langle C^2 \rangle_\infty$ in the ergodic case is helpful as a reference point. 
Parametrizing a generic final state of the system as $|\psi\rangle$ with $a = \cos(\psi)$,$ c = \sin(\psi)\cos(\theta)$, $ \alpha=\sin(\psi)\sin(\theta)\cos(\phi)$ and $\gamma =\sin(\psi)\sin(\theta)\sin(\phi) $, the measure on the state parameters becomes $d\mu = \sin (\theta ) \sin ^2(\psi ) d\theta d\phi d\psi$. 
In the ergodic regime, where all states are equally likely, the average squared concurrence is given by 
\begin{align}\label{eq:avC2}
    \langle C^2 \rangle &= \frac{1}{2 \pi^2}\int \int \int 4 \sin ^2(\theta ) \sin ^2(\psi ) Z(\theta, \psi,\phi)^2 d\mu  = \frac{1}{3},
\end{align}
where $Z(\theta, \psi,\phi) = \cos (\psi ) \sin (\phi )-\cos (\theta ) \sin (\psi ) \cos (\phi )$, which agrees with the no-measurement limit in the numerical simulations. (c.f. Fig.~\ref{fig:CvsGamma}).  Remarkably, increasing the noise strength under continuous monitoring has a non-monotonic effect on the concurrence. In particular, at any measurement strength, weak noise in the system increases the total steady-state entanglement, while larger noise strength induces a decrease in concurrence.

The effect can be understood by observing that, for low noise, increasing noise generically increases the probability of exploring larger parts of the Hilbert space (with higher entangled states). In contrast, larger local noise tends to induce large fluctuations of the local energy levels, which hinder the entangling effect of the two-qubit Hamiltonian.
This effect becomes inappreciable for vanishing measurements when the steady-state distribution tends to Eq.~\eqref{eq:avC2}, i.e. it is mainly independent of the noise strength. It also becomes gradually less pronounced for strong measurements where the disentangling effect of measurements dominates over the noise one.

\section{Conclusions}
In this work, we have derived a stochastic action for the dynamics of a coupled two-qubit system subject to a driving Hamiltonian, independent continuous Gaussian measurements and Gaussian noise.
The actions have been obtained from the CDJ formalism developed for monitored systems, which encompasses a path integral over all the state of the system's Hilbert space.  
We have combined the CDJ path integral with the standard path integral on Gaussian noise fluctuations and determined the effect of their interplay. 

First, we use the path integral to determine the most likely quantum trajectories and their associated concurrence.
This formalism allows us to analyse a subset associated with optimal quantum trajectories, showing that the optimal path captures some general features of the entanglement dynamics, like their oscillations and non-monotonicity.
However, it fails to capture quantitative features, e.g., oscillation amplitudes and max/min values. 
For the global optimal trajectory, the most likely path can also capture a transition from an oscillating behaviour at short times to an overdamped approach to a steady state as a function of the measurement strength. 
While such behaviour has a counterpart in the average dynamics, the optimal trajectory does not capture the correct steady-state dependence of the entanglement and misses the effect of the Gaussian noise.

Using the developed action formalism, we have obtained a diagrammatic approach for the entanglement dynamics that is valid for a short time, with weak measurements and noise. 
While the leading-order term in the expansion can capture the entanglement oscillations at short times with the appropriate frequency, the next non-trivial term showed a much better agreement of the oscillations' amplitudes. 
Notably, finite order expansion of the perturbation will generically lead to a vanishing steady state concurrence, which does not agree with the numerical simulation.
This hints at the need for some form of infinite resummation of the perturbation series to capture the long-time steady-state behaviour.

Finally, we obtained the steady-state entanglement from numerical simulations, which show a non-monotonic relationship between concurrence and noise strength in the system. 
These effects can be uniquely attributed to the entanglement along quantum trajectories as determined by measurement- and Hamiltonian-induced fluctuations and can serve as a first test of the ability to track non-linear quantum resources, like entanglement, along quantum trajectories.

\acknowledgments
We are grateful to H. Schomerus and Y.C. Leung for valuable discussions.

\appendix

\section{Interaction action}\label{ap:interaction}
We report the full expression for the interaction action $\tilde{S}_i$ in Eq.~\eqref{eq:stochasticAction} used in the diagrammatic approximation of $\langle C^2 \rangle$. The expression consists of 127 distinct terms, each associated with an interaction vertex. 

\onecolumngrid
\begin{align}\label{eq:interactionAction}
     &\tilde{S}_i = \frac{1}{\tau}\int dt \Bigg[ \frac{p_c^2 c^6}{2}+a p_a p_c c^5+\frac{p_c c^5}{2}+ p_c p_\alpha\alpha c^5+p_c p_\gamma \gamma c^5+\frac{a^2 p_a^2 c^4}{2 }+\frac{p_\alpha^2\alpha^2 c^4}{2 }+p_c^2 \gamma^2 c^4+\frac{p_\gamma^2 \gamma^2 c^4}{2 }+\frac{a p_a c^4}{2}+a p_a p_\alpha\alpha c^4 \nonumber \\ \nonumber &+\frac{p_\alpha\alpha c^4}{2 }+a p_a p_\gamma \gamma c^4+\frac{p_\gamma \gamma c^4}{2 }+p_\alpha p_\gamma\alpha \gamma c^4-p_c^2 c^4+2 p_c p_\gamma \gamma^3 c^3+2 a p_a p_c \gamma^2 c^3+p_c \gamma^2 c^3+2 p_c p_\alpha\alpha \gamma^2 c^3-a p_a p_c c^3-p_c c^3\\ \nonumber &-p_c p_\alpha\alpha c^3-2 p_c p_\gamma \gamma c^3+\frac{p_c^2\alpha^4 c^2}{2 }+p_c^2 \gamma^4 c^2+p_\gamma^2 \gamma^4 c^2+2 a p_a p_\gamma \gamma^3 c^2+p_\gamma \gamma^3 c^2+2 p_\alpha p_\gamma\alpha \gamma^3 c^2+\frac{1}{2} \Gamma  p_a^2 c^2+\frac{p_c^2 c^2}{2}+\frac{1}{2} \Gamma  p_\gamma^2 c^2\\ \nonumber &+a^2 p_a^2 \gamma^2 c^2+p_c^2\alpha^2 \gamma^2 c^2+p_\alpha^2\alpha^2 \gamma^2 c^2+a p_a \gamma^2 c^2+2 a p_a p_\alpha\alpha \gamma^2 c^2+p_\alpha\alpha \gamma^2 c^2-p_c^2 \gamma^2 c^2-p_\gamma^2 \gamma^2 c^2-a p_a p_\gamma \gamma c^2-p_\gamma \gamma c^2\\ \nonumber &-p_\alpha p_\gamma\alpha \gamma c^2+p_c p_\alpha\alpha^5 c+2 p_c p_\gamma \gamma^5 c+a p_a p_c\alpha^4 c+\frac{p_c\alpha^4 c}{2  }+2 a p_a p_c \gamma^4 c+p_c \gamma^4 c+2 p_c p_\alpha\alpha \gamma^4 c+2 p_c p_\gamma\alpha^2 \gamma^3 c+2 p_c p_\alpha\alpha^3 \gamma^2 c\\ \nonumber &+2 a p_a p_c\alpha^2 \gamma^2 c+p_c\alpha^2 \gamma^2 c-\Gamma  a p_a p_c c+\Gamma  a p_\alpha p_\gamma c-2 \Gamma  p_a p_\gamma\alpha c+p_c p_\gamma\alpha^4 \gamma c+\Gamma  p_a p_\alpha \gamma c-\Gamma  p_c p_\gamma \gamma c +p_c p_\gamma \gamma c-p_c p_\alpha\alpha^3 c\\ \nonumber &-3 p_c p_\gamma \gamma^3 c-a p_a p_c \gamma^2 c-p_c \gamma^2 c-2 p_c p_\alpha\alpha \gamma^2 c-p_c p_\gamma\alpha^2 \gamma c+\frac{p_\alpha^2\alpha^6}{2 }+p_\gamma^2 \gamma^6+a p_a p_\alpha\alpha^5+\frac{p_\alpha\alpha^5}{2 }+2 a p_a p_\gamma \gamma^5+p_\gamma \gamma^5\\ \nonumber &+2 p_\alpha p_\gamma\alpha \gamma^5+\frac{a^2 p_a^2\alpha^4}{2 }+\frac{a p_a\alpha^4}{2 }+a^2 p_a^2 \gamma^4+p_\alpha^2\alpha^2 \gamma^4+p_\gamma^2\alpha^2 \gamma^4+a p_a \gamma^4+2 a p_a p_\alpha\alpha \gamma^4+p_\alpha\alpha \gamma^4+2 p_\alpha p_\gamma\alpha^3 \gamma^3\\ \nonumber &+2 a p_a p_\gamma\alpha^2 \gamma^3+p_\gamma\alpha^2 \gamma^3+\frac{1}{2} \Gamma  a^2 p_c^2+\frac{1}{2} \Gamma  a^2 p_\alpha^2+\frac{1}{2} \Gamma  p_a^2\alpha^2+\frac{p_\alpha^2\alpha^2}{2  }+\frac{1}{2} \Gamma  p_\gamma^2\alpha^2+p_\alpha^2\alpha^4 \gamma^2+\frac{p_\gamma^2\alpha^4 \gamma^2}{2 }+2 a p_a p_\alpha\alpha^3 \gamma^2\\ \nonumber &+p_\alpha\alpha^3 \gamma^2+\frac{1}{2} \Gamma  p_c^2 \gamma^2+\frac{1}{2} \Gamma  p_\alpha^2 \gamma^2+p_\gamma^2 \gamma^2+a^2 p_a^2\alpha^2 \gamma^2+a p_a\alpha^2 \gamma^2 -\Gamma  a p_a p_\alpha\alpha+\Gamma  a p_c p_\gamma\alpha+p_\alpha p_\gamma\alpha^5 \gamma+a p_a p_\gamma\alpha^4 \gamma\\ \nonumber &+\frac{p_\gamma\alpha^4 \gamma}{2 }-2 \Gamma  a p_c p_\alpha \gamma+\Gamma  p_a p_c\alpha \gamma-\Gamma  p_\alpha p_\gamma\alpha \gamma+p_\alpha p_\gamma\alpha \gamma-p_\alpha^2\alpha^4 -2 p_\gamma^2 \gamma^4-a p_a p_\alpha\alpha^3-p_\alpha\alpha^3-2 a p_a p_\gamma \gamma^3-2 p_\gamma \gamma^3 \\  &-3 p_\alpha p_\gamma\alpha \gamma^3-p_\alpha^2\alpha^2 \gamma^2-p_\gamma^2\alpha^2 \gamma^2-a p_a p_\alpha\alpha \gamma^2-p_\alpha\alpha \gamma^2 -2 p_\alpha p_\gamma\alpha^3 \gamma-a p_a p_\gamma\alpha^2 \gamma-p_\gamma\alpha^2 \gamma \Bigg]
\end{align}
\twocolumngrid

\bibliographystyle{unsrt.bst}
\bibliography{apssamp}

\providecommand{\noopsort}[1]{}\providecommand{\singleletter}[1]{#1}%
\begin{thebibliography}{10}

\bibitem{nielsen2000quantum}
M.A. Nielsen and I.L. Chuang.
\newblock {\em Quantum Computation and Quantum Information}.
\newblock Cambridge Series on Information and the Natural Sciences. Cambridge University Press, 2000.

\bibitem{zeng2018quantum}
Bei Zeng, Xie Chen, Duan-Lu Zhou, and Xiao-Gang Wen.
\newblock Quantum information meets quantum matter -- from quantum entanglement to topological phase in many-body systems, 2018.

\bibitem{PhysRev.109.1492}
P.~W. Anderson.
\newblock Absence of diffusion in certain random lattices.
\newblock {\em Phys. Rev.}, 109:1492--1505, Mar 1958.

\bibitem{PhysRevLett.78.2803}
Boris~L. Altshuler, Yuval Gefen, Alex Kamenev, and Leonid~S. Levitov.
\newblock Quasiparticle lifetime in a finite system: A nonperturbative approach.
\newblock {\em Phys. Rev. Lett.}, 78:2803--2806, Apr 1997.

\bibitem{BASKO20061126}
D.M. Basko, I.L. Aleiner, and B.L. Altshuler.
\newblock Metal–insulator transition in a weakly interacting many-electron system with localized single-particle states.
\newblock {\em Annals of Physics}, 321(5):1126--1205, 2006.

\bibitem{ABANIN2021168415}
D.A. Abanin, J.H. Bardarson, G.~{De Tomasi}, S.~Gopalakrishnan, V.~Khemani, S.A. Parameswaran, F.~Pollmann, A.C. Potter, M.~Serbyn, and R.~Vasseur.
\newblock Distinguishing localization from chaos: Challenges in finite-size systems.
\newblock {\em Annals of Physics}, 427:168415, 2021.

\bibitem{RevModPhys.91.021001}
Dmitry~A. Abanin, Ehud Altman, Immanuel Bloch, and Maksym Serbyn.
\newblock Colloquium: Many-body localization, thermalization, and entanglement.
\newblock {\em Rev. Mod. Phys.}, 91:021001, May 2019.

\bibitem{PhysRevLett.96.110404}
Alexei Kitaev and John Preskill.
\newblock Topological entanglement entropy.
\newblock {\em Phys. Rev. Lett.}, 96:110404, Mar 2006.

\bibitem{Page_1993}
Don~N. Page.
\newblock Average entropy of a subsystem.
\newblock {\em Physical Review Letters}, 71(9):1291–1294, August 1993.

\bibitem{PhysRevB.98.205136}
Yaodong Li, Xiao Chen, and Matthew P.~A. Fisher.
\newblock Quantum zeno effect and the many-body entanglement transition.
\newblock {\em Phys. Rev. B}, 98:205136, Nov 2018.

\bibitem{PhysRevB.99.224307}
Amos Chan, Rahul~M. Nandkishore, Michael Pretko, and Graeme Smith.
\newblock Unitary-projective entanglement dynamics.
\newblock {\em Phys. Rev. B}, 99:224307, Jun 2019.

\bibitem{PhysRevX.9.031009}
Brian Skinner, Jonathan Ruhman, and Adam Nahum.
\newblock Measurement-induced phase transitions in the dynamics of entanglement.
\newblock {\em Phys. Rev. X}, 9:031009, Jul 2019.

\bibitem{PhysRevX.7.031016}
Adam Nahum, Jonathan Ruhman, Sagar Vijay, and Jeongwan Haah.
\newblock Quantum entanglement growth under random unitary dynamics.
\newblock {\em Phys. Rev. X}, 7:031016, Jul 2017.

\bibitem{PhysRevB.100.064204}
M.~Szyniszewski, A.~Romito, and H.~Schomerus.
\newblock Entanglement transition from variable-strength weak measurements.
\newblock {\em Phys. Rev. B}, 100:064204, Aug 2019.

\bibitem{Fisher_2023}
Matthew~P.A. Fisher, Vedika Khemani, Adam Nahum, and Sagar Vijay.
\newblock Random quantum circuits.
\newblock {\em Annual Review of Condensed Matter Physics}, 14(1):335–379, March 2023.

\bibitem{Noel_2022}
Crystal Noel, Pradeep Niroula, Daiwei Zhu, Andrew Risinger, Laird Egan, Debopriyo Biswas, Marko Cetina, Alexey~V. Gorshkov, Michael~J. Gullans, David~A. Huse, and Christopher Monroe.
\newblock Measurement-induced quantum phases realized in a trapped-ion quantum computer.
\newblock {\em Nature Physics}, 18(7):760–764, June 2022.

\bibitem{Koh_2023}
Jin~Ming Koh, Shi-Ning Sun, Mario Motta, and Austin~J. Minnich.
\newblock Measurement-induced entanglement phase transition on a superconducting quantum processor with mid-circuit readout.
\newblock {\em Nature Physics}, 19(9):1314–1319, June 2023.

\bibitem{2023}
Google AI and Collaborators.
\newblock Measurement-induced entanglement and teleportation on a noisy quantum processor.
\newblock {\em Nature}, 622(7983):481–486, October 2023.

\bibitem{PhysRevB.67.241305}
Rusko Ruskov and Alexander~N. Korotkov.
\newblock Entanglement of solid-state qubits by measurement.
\newblock {\em Phys. Rev. B}, 67:241305, Jun 2003.

\bibitem{Moehring_2007}
D~Moehring, Peter Maunz, S~Olmschenk, Kelly Younge, Dzmitry Matsukevich, L.-M Duan, and C~Monroe.
\newblock Entanglement of single-atom quantum bits at a distance.
\newblock {\em Nature}, 449:68--71, 10 2007.

\bibitem{Hutchison_2009}
Chantal~L. Hutchison, J.~M. Gambetta, Alexandre Blais, and F.~K. Wilhelm.
\newblock Quantum trajectory equation for multiple qubits in circuit qed: Generating entanglement by measurementthis paper was presented at the theory canada 4 conference, held at centre de recherches mathématiques, montréal, québec, canada on 4–7 june 2008.
\newblock {\em Canadian Journal of Physics}, 87(3):225–231, March 2009.

\bibitem{PhysRevA.81.040301}
Kevin Lalumi\`ere, J.~M. Gambetta, and Alexandre Blais.
\newblock Tunable joint measurements in the dispersive regime of cavity qed.
\newblock {\em Phys. Rev. A}, 81:040301, Apr 2010.

\bibitem{Bernien_2013}
H.~Bernien, B.~Hensen, W.~Pfaff, G.~Koolstra, M.~S. Blok, L.~Robledo, T.~H. Taminiau, M.~Markham, D.~J. Twitchen, L.~Childress, and R.~Hanson.
\newblock Heralded entanglement between solid-state qubits separated by three metres.
\newblock {\em Nature}, 497(7447):86–90, April 2013.

\bibitem{PhysRevA.83.022311}
Eduardo Mascarenhas, Daniel Cavalcanti, Vlatko Vedral, and Marcelo Fran\ifmmode \mbox{\c{c}}\else~\c{c}\fi{}a Santos.
\newblock Physically realizable entanglement by local continuous measurements.
\newblock {\em Phys. Rev. A}, 83:022311, Feb 2011.

\bibitem{PhysRevLett.112.170501}
N.~Roch, M.~E. Schwartz, F.~Motzoi, C.~Macklin, R.~Vijay, A.~W. Eddins, A.~N. Korotkov, K.~B. Whaley, M.~Sarovar, and I.~Siddiqi.
\newblock Observation of measurement-induced entanglement and quantum trajectories of remote superconducting qubits.
\newblock {\em Phys. Rev. Lett.}, 112:170501, Apr 2014.

\bibitem{PhysRevX.6.041052}
Areeya Chantasri, Mollie~E. Kimchi-Schwartz, Nicolas Roch, Irfan Siddiqi, and Andrew~N. Jordan.
\newblock Quantum trajectories and their statistics for remotely entangled quantum bits.
\newblock {\em Phys. Rev. X}, 6:041052, Dec 2016.

\bibitem{PhysRevResearch.2.033512}
Kyrylo Snizhko, Parveen Kumar, and Alessandro Romito.
\newblock Quantum zeno effect appears in stages.
\newblock {\em Phys. Rev. Res.}, 2:033512, Sep 2020.

\bibitem{PhysRevA.91.012118}
Pierre Rouchon and Jason~F. Ralph.
\newblock Efficient quantum filtering for quantum feedback control.
\newblock {\em Phys. Rev. A}, 91:012118, Jan 2015.

\bibitem{PhysRevA.102.022612}
Yuxiao Jiang, Xiyue Wang, Leigh Martin, and K.~Birgitta Whaley.
\newblock Optimality of feedback control for qubit purification under inefficient measurement.
\newblock {\em Phys. Rev. A}, 102:022612, Aug 2020.

\bibitem{Kalsi_2022}
T~Kalsi, A~Romito, and H~Schomerus.
\newblock Three-fold way of entanglement dynamics in monitored quantum circuits.
\newblock {\em Journal of Physics A: Mathematical and Theoretical}, 55(26):264009, June 2022.

\bibitem{PhysRevB.105.144202}
T.~Boorman, M.~Szyniszewski, H.~Schomerus, and A.~Romito.
\newblock Diagnostics of entanglement dynamics in noisy and disordered spin chains via the measurement-induced steady-state entanglement transition.
\newblock {\em Phys. Rev. B}, 105:144202, Apr 2022.

\bibitem{Schomerus_2022}
Henning Schomerus.
\newblock Noisy monitored quantum dynamics of ergodic multi-qubit systems.
\newblock {\em Journal of Physics A: Mathematical and Theoretical}, 55(21):214001, May 2022.

\bibitem{PhysRevA.88.042110}
A.~Chantasri, J.~Dressel, and A.~N. Jordan.
\newblock Action principle for continuous quantum measurement.
\newblock {\em Phys. Rev. A}, 88:042110, Oct 2013.

\bibitem{PhysRevA.92.032125}
Areeya Chantasri and Andrew~N. Jordan.
\newblock Stochastic path-integral formalism for continuous quantum measurement.
\newblock {\em Phys. Rev. A}, 92:032125, Sep 2015.

\bibitem{Murch_2013}
K.~W. Murch, S.~J. Weber, C.~Macklin, and I.~Siddiqi.
\newblock Observing single quantum trajectories of a superconducting quantum bit.
\newblock {\em Nature}, 502(7470):211--214, oct 2013.

\bibitem{PhysRevA.95.042126}
Philippe Lewalle, Areeya Chantasri, and Andrew~N. Jordan.
\newblock Prediction and characterization of multiple extremal paths in continuously monitored qubits.
\newblock {\em Phys. Rev. A}, 95:042126, Apr 2017.

\bibitem{PhysRevA.98.012141}
Philippe Lewalle, John Steinmetz, and Andrew~N. Jordan.
\newblock Chaos in continuously monitored quantum systems: An optimal-path approach.
\newblock {\em Phys. Rev. A}, 98:012141, Jul 2018.

\bibitem{PhysRevA.97.012118}
Areeya Chantasri, Juan Atalaya, Shay Hacohen-Gourgy, Leigh~S. Martin, Irfan Siddiqi, and Andrew~N. Jordan.
\newblock Simultaneous continuous measurement of noncommuting observables: Quantum state correlations.
\newblock {\em Phys. Rev. A}, 97:012118, Jan 2018.

\bibitem{PRXQuantum.3.010327}
Tathagata Karmakar, Philippe Lewalle, and Andrew~N. Jordan.
\newblock Stochastic path-integral analysis of the continuously monitored quantum harmonic oscillator.
\newblock {\em PRX Quantum}, 3:010327, Feb 2022.

\bibitem{jacobs_2014}
Kurt Jacobs.
\newblock {\em Quantum Measurement Theory and its Applications}.
\newblock Cambridge University Press, 2014.

\bibitem{Jacobs_2006}
Kurt Jacobs and Daniel~A. Steck.
\newblock A straightforward introduction to continuous quantum measurement.
\newblock {\em Contemporary Physics}, 47(5):279--303, sep 2006.

\bibitem{kleinert2009path}
H.~Kleinert.
\newblock {\em Path Integrals In Quantum Mechanics, Statistics, Polymer Physics, And Financial Markets (5th Edition)}.
\newblock World Scientific Publishing Company, 2009.

\bibitem{altland_simons_2010}
Alexander Altland and Ben~D. Simons.
\newblock {\em Condensed Matter Field Theory}.
\newblock Cambridge University Press, 2 edition, 2010.

\bibitem{chow2012path}
Carson~C. Chow and Michael~A. Buice.
\newblock Path integral methods for stochastic differential equations.
\newblock {\em The Journal of Mathematical Neuroscience (JMN)}, 5:8, 2015.

\bibitem{wio2013path}
H.S. Wio.
\newblock {\em Path Integrals for Stochastic Processes: An Introduction}.
\newblock World Scientific, 2013.

\end{thebibliography}

\end{document}